\documentclass[pdflatex, sn-mathphys-num]{sn-jnl}

\usepackage{graphicx}
\usepackage{multirow}
\usepackage{amsmath,amssymb,amsfonts}
\usepackage{amsthm}
\usepackage{mathrsfs}
\usepackage[title]{appendix}
\usepackage{xcolor}
\usepackage{textcomp}
\usepackage{manyfoot}
\usepackage{listings}
\usepackage{physics}
\usepackage{subcaption}

\title{Perfect fluid dark matter: a viability test with galaxy rotation curves}
\author*[1]{\fnm{Jan} \sur{Kuncewicz}}\email{kuncewiczjan@gmail.com}
\affil[1]{
    \orgdiv{Institute of Physics},
    \orgname{Maria Curie-Sk\l{}odowska University},
    \orgaddress{\street{pl.~Marii~Curie-Sk\l{}odowskiej~1},
    \city{Lublin}, \postcode{20-031}, \country{Poland}}
}

\usepackage{amsmath,amssymb,amsfonts}
\usepackage{amsthm}
\usepackage{physics}
\usepackage{subcaption}

\abstract{
The anomalous rotation curves of galaxies provide compelling evidence for dark
matter, yet its fundamental nature and distribution remain key unresolved issues
in astrophysics. In this work, we investigate a dark matter model derived from
first principles within General Relativity, treating the halo as a perfect fluid
with a specific anisotropic equation of state characterized by a single
parameter. This framework yields two families of static, spherically
symmetric solutions: a Power-Law metric and a Logarithmic metric. As an initial
viability test, we fit the model's derived circular velocity profiles to the
dark matter contributions of representative galaxies from the SPARC database.
Our analysis reveals that the two solutions effectively describe different
regions of the halo: the Logarithmic form accurately models the large-radius
behavior, while the Power-Law form successfully reproduces the inner rotation
curve. Notably, the model consistently favors a shallow central density profile,
aligning with cored halo models and providing a fit for galaxies with
a gradual rise in velocity. We conclude that this simple, analytically-derived
fluid model provides a compelling and physically-motivated framework for
describing galactic rotation curves, warranting a more exhaustive study across a
larger sample of galaxies.
}

\keywords{Dark Matter, Black Holes, General Relativity, Galaxy Rotation Curve, Perfect Fluid Dark Matter}

\begin{document}
\maketitle
\section{Introduction}
The existence of dark matter is one of the most significant and well-established
puzzles in modern cosmology and astrophysics. A wealth of observational
evidence, including the anomalous rotation curves of galaxies
\cite{rubin1970rotation, corbelli2000extended}, anisotropies in the cosmic
microwave background radiation \cite{madhavacheril2014current}, and
gravitational lensing phenomena \cite{clowe2006direct}, points towards the
existence of a non-luminous matter component that dominates the gravitational
dynamics of the universe. In particular, the predictions of Newtonian gravity
and General Relativity, when applied to the visible baryonic matter alone,
cannot be reconciled with the observed flat asymptotic behavior of galactic
rotation curves \cite{rubin1970rotation}.

This discrepancy has catalyzed two principal lines of inquiry. The first
proposes modifications to the established theories of gravity, such as Modified
Newtonian Dynamics (MOND) and its relativistic extensions
\cite{milgrom1983modification, bekenstein2004relativistic, bekenstein1984does,
capozziello2012dark, li2012haloes}. The second, and more widely accepted
paradigm, involves the introduction of new, non-baryonic matter components that
interact weakly, if at all, with the particles of the Standard Model
\cite{gervais1971field, wess1974supergauge, peccei1977constraints}. These
hypothetical particles, often motivated by frameworks like supersymmetry
\cite{jungman1996supersymmetric}, are collectively termed dark matter.

On the phenomenological front, significant effort has been dedicated to modeling
the spatial distribution of dark matter within galaxies. These models generally
fall into two categories. \textit{Cuspy} profiles, which predict a steeply rising
density towards the galactic center ($\rho\propto 1/r^\gamma$), are favored by
N-body simulations of cold dark matter \cite{navarro1995simulations,
navarro1997universal, moore1999cold, fukushige2001structure}. In contrast,
\textit{cored} profiles, characterized by a nearly constant density core, appear to be
more consistent with observations of dwarf and low-surface-brightness galaxies
\cite{burkert1995structure, salucci2000dark, brownstein2006galaxy}. This tension
between simulation and observation, known as the \textit{cusp-vs-core} problem, remains
a key challenge for dark matter models \cite{moore1994evidence, oh2015high}.

Within the context of particle-based solutions, a variety of analytic halo models
have been proposed. Some notable examples employ scalar fields to describe the
dark matter halo, such as the Brans-Dicke massless scalar field used by Fay
\cite{fay2004scalar} or the minimally coupled scalar field with a potential
investigated by Matos, Guzmán, and Nuñez \cite{matos2000spherical}. Following a
similar avenue, this paper investigates a dark matter model described by a
perfect fluid with a specific, barotropic equation of state. This approach,
framed within General Relativity, leads to static, spherically symmetric
solutions of the Einstein field equations characterized by a single parameter,
$\epsilon$. Such solutions have been previously explored in various contexts
\cite{salgado2003simple, dymnikova2002cosmological, giambo2002anisotropic,
kiselev2003quintessence}.

The primary objective of this work is to assess the viability of this simple,
analytically-derived model by confronting it with observational data. We derive
the tangential velocities for circular orbits within this spacetime and compare
them to the comprehensive SPARC (Spitzer Photometry and Accurate Rotation
Curves) database \cite{lelli2016sparc}. We demonstrate that our model, utilizing
two distinct functional forms depending on the value of $\epsilon$, can
effectively describe the entire profile of a galaxy's rotation curve. Notably,
the resulting dark matter density profiles are more analogous to cored models.
The strength of this approach lies in providing a good fit to observational data
from a simple theoretical foundation, without imposing empirically motivated
density profiles from the outset.

\section{Theoretical Framework and Derivation of Orbital Velocities}
The theoretical basis for our analysis is a static and spherically symmetric
spacetime, whose geometry is sourced by a central baryonic mass and a
surrounding dark matter component. This dark matter is modeled as a fluid with a
specific anisotropic pressure profile. Following the approach in
\cite{kuncewicz2025impacts, li2012galactic, salgado2003simple}, we consider a
diagonal energy-momentum tensor whose components satisfy the relation:
\begin{equation}
	T^{\theta}_\theta = T^\phi_\phi = T^t_t(1-\epsilon).
\end{equation}
Here, $\epsilon$ is a dimensionless constant parameterizing the fluid's equation
of state, linking the tangential pressure ($p_t = T^\theta_\theta$) to the
energy density ($\rho = -T^t_t$).

We adopt the standard metric ansatz for a static, spherically symmetric
spacetime:
\begin{equation}
	\dd s^2 = -f(r)\dd t^2 + f(r)^{-1}\dd r^2 + r^2\dd\theta^2 + r^2\sin^2\theta\dd\phi^2.
\end{equation}
Solving the Einstein field equations with this metric and the matter source
described above yields a family of solutions for the metric function $f(r)$.
Depending on the value of $\epsilon$, two distinct functional forms emerge:
\begin{align}
    \label{eq:met-exp}
	f(r) &= 1 - \frac{r_s}{r} + \frac{r^{2(1-\epsilon)}}{r_\epsilon},\ \epsilon \neq \frac32,\\
	\label{eq:met-log}
	f(r) &= 1 - \frac{r_s}{r} + \frac{a}{r}\ln\qty(\frac{r}{\abs{a}}),\ \epsilon = \frac32,
\end{align}
where $r_s = 2M$ is the Schwarzschild radius corresponding to the central
baryonic mass, and $r_\epsilon$ and $a$ are integration constants related to the
dark matter distribution. For clarity, we will refer to the solution
\eqref{eq:met-exp} as the \textit{Power-Law metric} and to \eqref{eq:met-log} as
the \textit{Logarithmic metric}. It is noteworthy that for specific values of
$\epsilon$, this model can reproduce the form of other well-known solutions,
such as the Reissner-Nördstrom or de Sitter spacetimes, highlighting its
versatility \cite{li2012galactic, kuncewicz2025impacts}.

To determine the galactic rotation curves predicted by this model, we analyze
the motion of massive test particles in this spacetime. Due to the
metric's symmetries, two constants of motion exist for a test particle following
a geodesic: the specific energy, $\mathcal{E}$, and the specific angular
momentum, $\mathcal{L}$, given by
\begin{equation}
	\mathcal{E} = f(r)\dv{t}{\tau},\ \mathcal{L} = r^2\dv{\phi}{\tau},
\end{equation}
where $\tau$ is the proper time along the geodesic. The radial equation of
motion can be expressed as:
\begin{equation}
	\qty(\dv{r}{\tau})^2 = \mathcal{E}^2 - \qty(\frac{\mathcal{L}^2}{r^2} + 1)f(r).
\end{equation}
The term multiplying $f(r)$ is part of the effective potential, $V_{eff}$, which
governs the radial motion. As established in \cite{kuncewicz2025impacts}, stable
circular orbits are characterized by two conditions: the radial velocity must be
zero ($\dv*{r}{\tau} = 0$), and the orbit must reside at a minimum of the
effective potential ($\dv*{V_{eff}}{r} = 0$). Combining these conditions allows
for the derivation of the tangential velocity $v$ of a particle in a circular
orbit:
\begin{equation}
	v^2(r) = r^2\qty(\dv{\phi}{t})^2 = r^2\qty(\dv{\phi}{\tau}/ \dv{t}{\tau})^2 = \frac12 r f'(r).
\end{equation}
Applying this general result to our specific metric functions,
\eqref{eq:met-exp} and \eqref{eq:met-log}, we obtain the squared orbital
velocities:
\begin{align}
	v^2(r) &= \frac{r_s}{2r} + \frac{(1 - \epsilon) r^{2(1-\epsilon)}}{r_\epsilon},\\
	v^2(r) &= \frac{r_s}{2r} + \frac{a}{2r}\qty[1 - \ln\qty(\frac{r}{\abs{a}})].
\end{align}
The term $r_s/2r$ corresponds to the standard Newtonian and Schwarzschild
contribution from the central mass. Consequently, we isolate the additional
velocity component generated by the dark matter halo, which we denote as
$\Delta v^2(r)$:
\begin{align}
    \label{eq:vel-exp-1}
	\Delta v^2(r) &= \frac{(1 - \epsilon) r^{2(1-\epsilon)}}{r_\epsilon},\\
	\label{eq:vel-log}
	\Delta v^2(r) &= \frac{a}{2r}\qty[1 - \ln\qty(\frac{r}{\abs{a}})].
\end{align}
The constant $r_\epsilon$ in \eqref{eq:vel-exp-1} is problematic for analysis,
as its physical units must vary with the parameter $\epsilon$ to ensure
dimensional consistency. To address this and obtain a more physically intuitive
parameterization, we introduce a new constant, $\lambda$, with units of length,
through the substitution
\begin{equation}
    \lambda = r_\epsilon^{1/[2(1-\epsilon)]}.
    \label{eq:lambda-def}
\end{equation}
This recasts equation \eqref{eq:vel-exp-1} into the more tractable form:
\begin{equation}
	\Delta v^2(r) = (1 - \epsilon)\qty(\frac{r}{\lambda})^{2(1-\epsilon)}.
	\label{eq:vel-exp-2}
\end{equation}
For the dark matter component to produce an attractive gravitational effect,
consistent with the observed enhancement of rotation velocities, the term
$\Delta v^2(r)$ must be positive. This imposes the physical constraint $\epsilon < 1$
for the Power-Law model.

While the substitution involving $\lambda$ provides a parameter with consistent
physical units, an alternative approach often convenient for numerical analysis
is to render the constant dimensionless. By introducing a fiducial length scale,
which we take to be 1~kpc, Eq. \eqref{eq:vel-exp-1} can be expressed as:
\begin{equation}
	\Delta v^2(r) = \frac{(1-\epsilon)(r/1\text{ kpc})^{2(1-\epsilon)}}{\tilde{r}_\epsilon}.
\end{equation}
In this formulation, the parameter $\tilde{r}_\epsilon = r_\epsilon / (1 \text{
kpc})^{2(1-\epsilon)}$ is a dimensionless quantity that characterizes the
strength of the dark matter contribution.

\section{Parameter Estimation from Observational Data}
\begin{table*}[ht!]
    \centering
    \caption{Best-fit parameters for the Logarithmic and Power-Law dark matter
    models derived from SPARC data. The parameter $a$ is determined from the
    large-radius behavior using Eq. \eqref{eq:vel-log}. The parameter $\epsilon$
    and the characteristic length scale $\lambda$ are obtained by fitting Eq.
    \eqref{eq:vel-exp-2} to the full rotation curve data. Due to parameter
    degeneracy, only the order of magnitude for $\lambda$ is provided.
    References point to the original sources of the observational data.}
    \label{tab:calc}
    \begin{tabular}{c c c c c}
    \hline
    Name & $a$ [$10^{-6}$ kpc] & $\epsilon$ & $\log(\lambda/ \text{1 kpc})$ & Ref.\\\hline
    UGC11455 & $-2.84232 \pm 0.25235$ & $0.60467 \pm 0.00119$ & 9  & \cite{spekkens2006structure}\\
    UGC08490 & $-0.06706 \pm 0.00741$ & $0.83189 \pm 0.00030$ & 20 & \cite{swaters2009rotation, swaters2002westerbork}\\
    UGC08286 & $-0.06105 \pm 0.00338$ & $0.75103 \pm 0.00067$ & 14 & \cite{swaters2009rotation, swaters2002westerbork}\\
    UGC07603 & $-0.01795 \pm 0.00001$ & $0.58489 \pm 0.00151$ & 9  & \cite{swaters2009rotation, swaters2002westerbork}\\
    UGC05986 & $-0.11553 \pm 0.01668$ & $0.54180 \pm 0.00168$ & 8  & \cite{swaters2009rotation, swaters2002westerbork}\\
    UGC03205 & $-2.26572 \pm 0.00038$ & $0.90494 \pm 0.00011$ & 30 & \cite{noordermeer2007mass, noordermeer2005westerbork}\\
    UGC01281 & $-0.01587 \pm 0.00272$ & $0.31668 \pm 0.00182$ & 6  & \cite{de2002high, swaters2002westerbork}\\
    NGC6503  & $-0.34980 \pm 0.07337$ & $0.83716 \pm 0.00024$ & 20 & \cite{begeman1991extended, begeman1989hi}\\
    NGC4559  & $-0.25356 \pm 0.03533$ & $0.63006 \pm 0.00055$ & 10 & \cite{barbieri2005extra}\\
    NGC4157  & $-0.99114 \pm 0.19316$ & $0.63970 \pm 0.00070$ & 10 & \cite{verheijen2001ursa, sanders1998rotation}\\
    NGC3198  & $-1.06859 \pm 0.05149$ & $0.84022 \pm 0.00024$ & 20 & \cite{daigle2006halpha,begeman1991extended, begeman1989hi}\\
    NGC2998  & $-1.67715 \pm 0.06410$ & $0.84678 \pm 0.00019$ & 20 & \cite{sanders1996published, courteau1996evidence}\\
    NGC1090  & $-0.69497 \pm 0.02538$ & $0.75809 \pm 0.00070$ & 14 & \cite{gentile04cored}\\
    NGC0801  & $-2.46766 \pm 0.27683$ & $0.87946 \pm 0.00031$ & 25 & \cite{sanders1996published, courteau1996evidence}\\
    NGC0024  & $-0.15979 \pm 0.00935$ & $0.77620 \pm 0.00070$ & 15 & \cite{dicaire2008halpha, chemin2006hi}\\
\end{tabular}
\end{table*}
To test the viability of our theoretical models, we confront their predictions
with observational data. For this purpose, we utilize the Spitzer Photometry and
Accurate Rotation Curves (SPARC) database \cite{lelli2016sparc}, which provides
high-quality rotation curve data for a large sample of galaxies. We selected a
subset of 175 galaxies, chosen specifically for their well-resolved kinematics
and prominent, extended flat velocity profiles, which provide a clear signature
of dark matter dominance at large radii.

The first step in our analysis is to isolate the velocity contribution from the
dark matter halo ($v_{DM}$). This is achieved by subtracting the velocity
contributions of the visible baryonic components from the observed rotation
curve ($v_{\text{obs}}$). Following the methodology outlined in
\cite{li2020comprehensive}, we calculate the squared dark matter velocity as:
\begin{equation}
	v_{DM}^2 = v_{\text{obs}}^2 - \Upsilon_{\text{disk}}v_{\text{disk}}^2
	- \Upsilon_{\text{bul}}v_{\text{bul}}^2 - v^2_{\text{gas}},
\end{equation}
where $v_{\text{disk}}$, $v_{\text{bul}}$, and $v_{\text{gas}}$ represent the
rotational velocities supported by the stellar disk, the stellar bulge, and the
gas component, respectively. The terms $\Upsilon_{\text{disk}}$ and
$\Upsilon_{\text{bul}}$ are the mass-to-light ratios for the disk and bulge. For
consistency with recent studies, we adopt the physically motivated values of
$\Upsilon_{\text{disk}} = 0.5$ and $\Upsilon_{\text{bul}} = 0.7$
\cite{li2020comprehensive, mcgaugh2016radial}.

Our fitting procedure involves a two-step process corresponding to our two
models. First, we estimate the parameter $a$ for the Logarithmic metric
\eqref{eq:vel-log}. This model predicts an asymptotic velocity fall-off of $v
\propto \sqrt{\ln{r}/r}$, a behavior consistent with the large-radius
predictions of several established dark matter models \cite{sofue2020rotation}.
We therefore determine $a$ by fitting Eq. \eqref{eq:vel-log} to the outermost
data points of the derived $v_{DM}$ profile for each galaxy.

Next, we analyze the Power-Law model \eqref{eq:vel-exp-2} to determine the
parameter $\epsilon$. This is accomplished by performing a non-linear
least-squares fit of the model to the entire $v_{DM}(r)$ profile. During this
procedure, a notable feature of the parameter $\lambda$ emerged: for all
galaxies, the best-fit value of $\lambda$ was found to be orders of magnitude
larger than the maximum radial extent of the observational data. This leads to a
parameter degeneracy, as for any radius $r$ within the galaxy ($r \ll \lambda$),
the model's prediction becomes extremely insensitive to the precise value of
$\lambda$. Mathematically, the gradient of the velocity function with respect to
$\lambda$ approaches zero: $\pdv*{v_{DM}^2}{\lambda} \propto
\lambda^{-(2(1-\epsilon)+1)} \approx 0$. Consequently, while the fit robustly
constrains $\epsilon$, it is not possible to determine $\lambda$ with any
meaningful precision. For this reason, we only report its estimated order of
magnitude.

The results of this analysis for both models are presented in Table
\ref{tab:calc}. The table lists the fitted values for $a$ and $\epsilon$ with
their statistical uncertainties, alongside the order of magnitude for the scale
length $\lambda$.

\section{Discussion}
The results presented in the previous section demonstrate that the perfect fluid
dark matter model, despite its theoretical simplicity, can effectively reproduce
key features of observed galactic rotation curves. The two distinct solutions,
the Logarithmic and the Power-Law metrics, appear to describe different radial
domains of the dark matter halo, suggesting they may act as complementary
components of a more unified description.

The Logarithmic metric (corresponding to $\epsilon = 3/2$) proves to be
particularly well-suited for describing the outer regions of the galaxies. The
velocity profile it generates, $\Delta v^2(r) \propto (1/r)(1 - \ln(r/|a|))$,
exhibits an asymptotic fall-off consistent with the $v \propto \sqrt{\ln r/r}$
behavior predicted by many widely-used empirical halo models, both cored and
cuspy \cite{burkert1995structure, salucci2000dark, navarro1997universal,
sofue2020rotation}. This reinforces the validity of this solution in the
large-$r$ limit. Furthermore, the parameter $a$ has a direct physical
interpretation. As per the convention in \cite{salgado2003simple}, the energy
density of the fluid is given by $\rho = -T^t_t = -a/r^3$. Our fitting procedure
consistently yields negative values for $a$ (Table \ref{tab:calc}), which
corresponds to a positive, physically sensible dark matter density
that decreases with radius. We observe no simple correlation between the value
of $a$ and global galaxy properties such as mass or size, suggesting that the
dark matter distribution is highly dependent on the individual characteristics
and formation history of each galaxy.

\begin{figure}[ht!]
    \centering
    \begin{subfigure}{.7\textwidth}
        \centering
        \includegraphics[width=\textwidth]{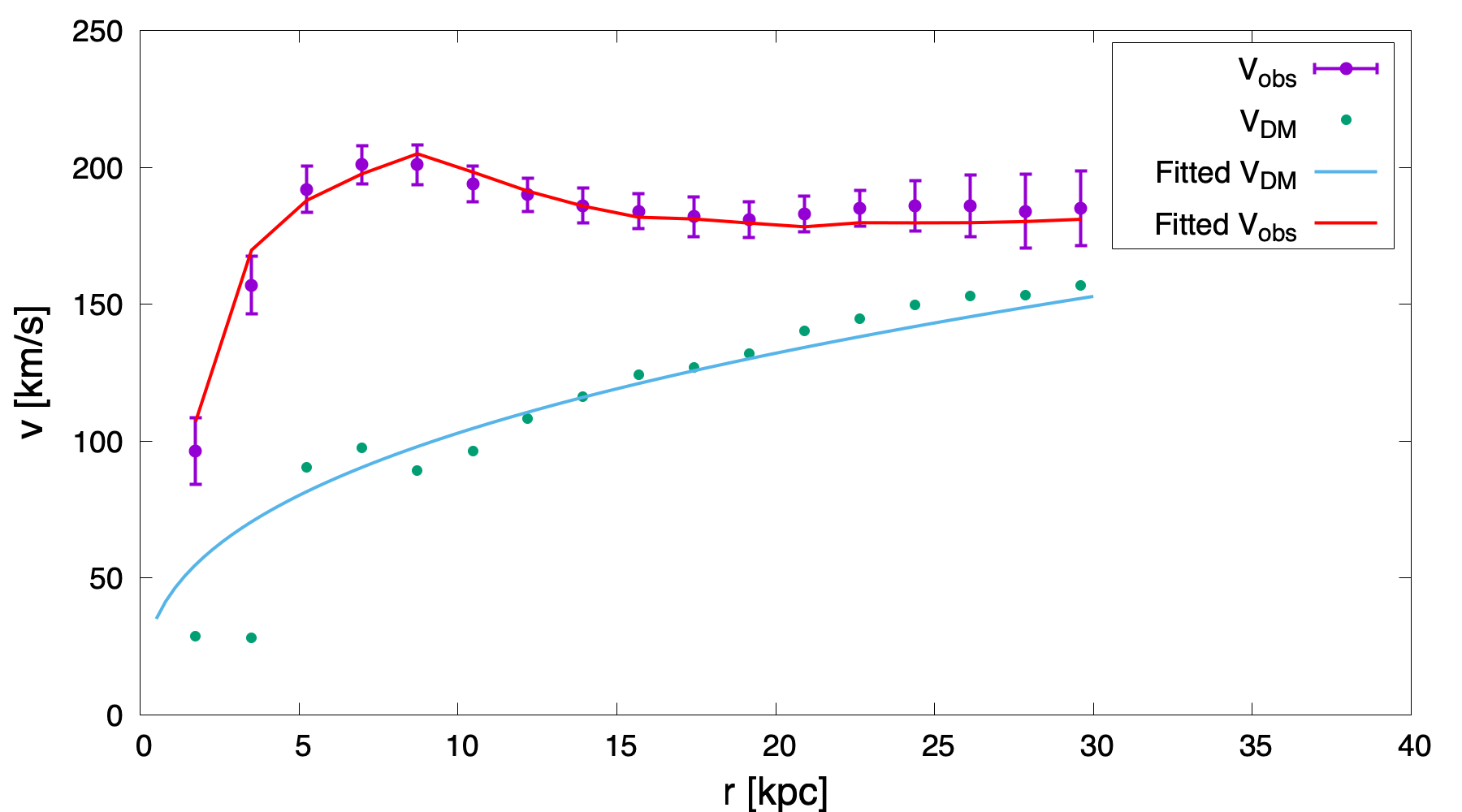}
    \end{subfigure}
    \begin{subfigure}{.7\textwidth}
        \centering
        \includegraphics[width=\textwidth]{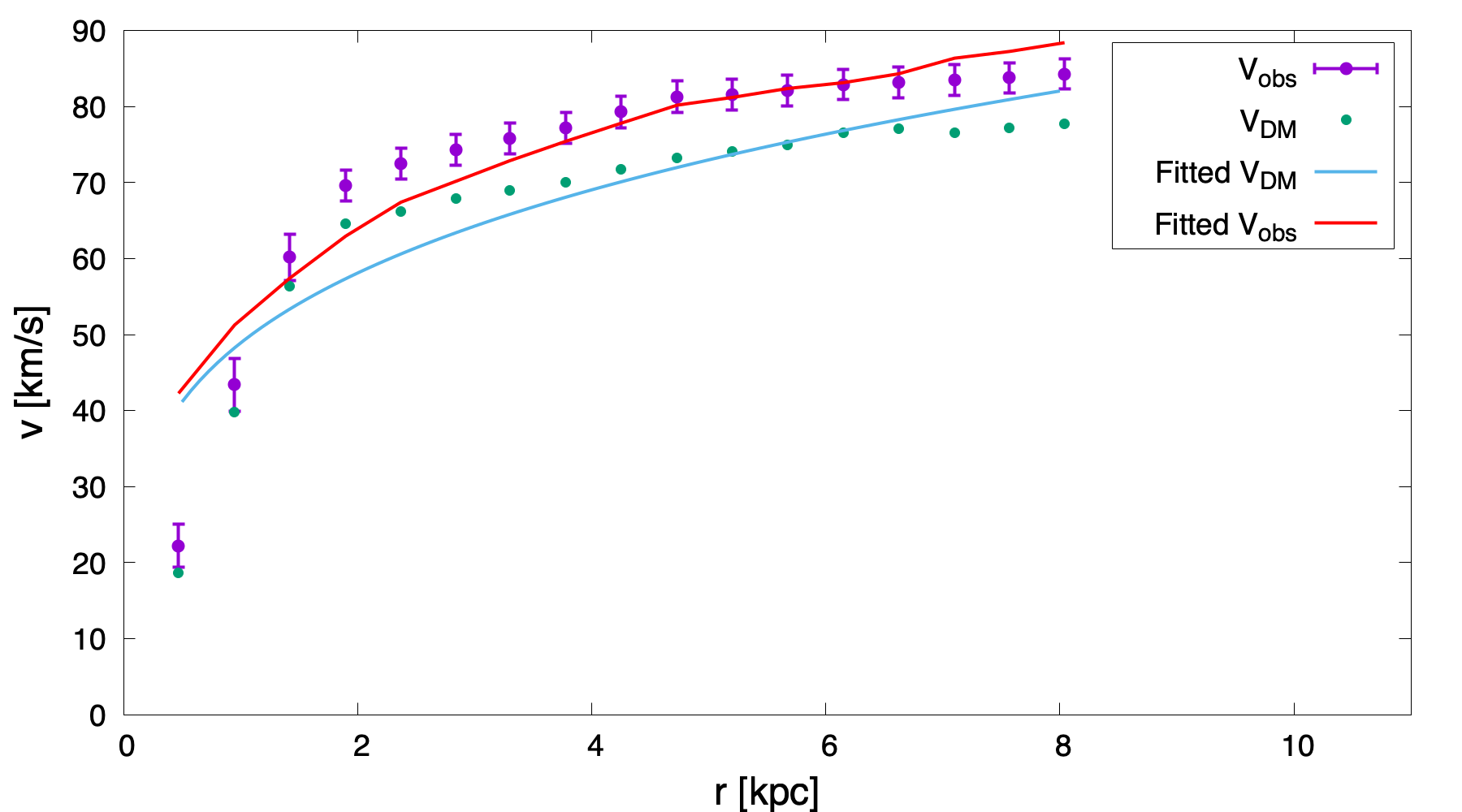}
    \end{subfigure}
    \caption{Comparison of the Power-Law model fit to two galaxies from the
    SPARC sample. The model provides a significantly better fit for galaxies
    with a gradual velocity rise (top, NGC4157, $\chi_\nu^2 = 0.251$) than for
    those with a steep inner gradient (bottom, UGC08286, $\chi_\nu^2 = 6.088$).}
    \label{fig:rot-curve}
\end{figure}
In contrast, the Power-Law model (for $\epsilon < 1$) is more adept at
describing the inner and intermediate regions of the rotation curve, where the
velocity rises and flattens. The quality of the fit, however, is strongly
dependent on the specific morphology of the galaxy's rotation curve. As
illustrated in Figure \ref{fig:rot-curve}, the model provides an excellent fit
for galaxies with a gradual, slow-rising velocity profile (e.g., NGC4157, with
$\chi_\nu^2 = 0.251$). It performs less well for galaxies that exhibit a very
steep rise in velocity near the galactic center (e.g., UGC08286, with $\chi_\nu^2
= 6.088$). This behavior indicates that our model inherently favors a shallower
density profile in the central region, making it more consistent with cored dark
matter models \cite{burkert1995structure} than with the steeply rising cuspy
profiles predicted by N-body simulations \cite{navarro1997universal}.

\begin{figure}[t]
	\centering
	\includegraphics[width=0.7\textwidth]{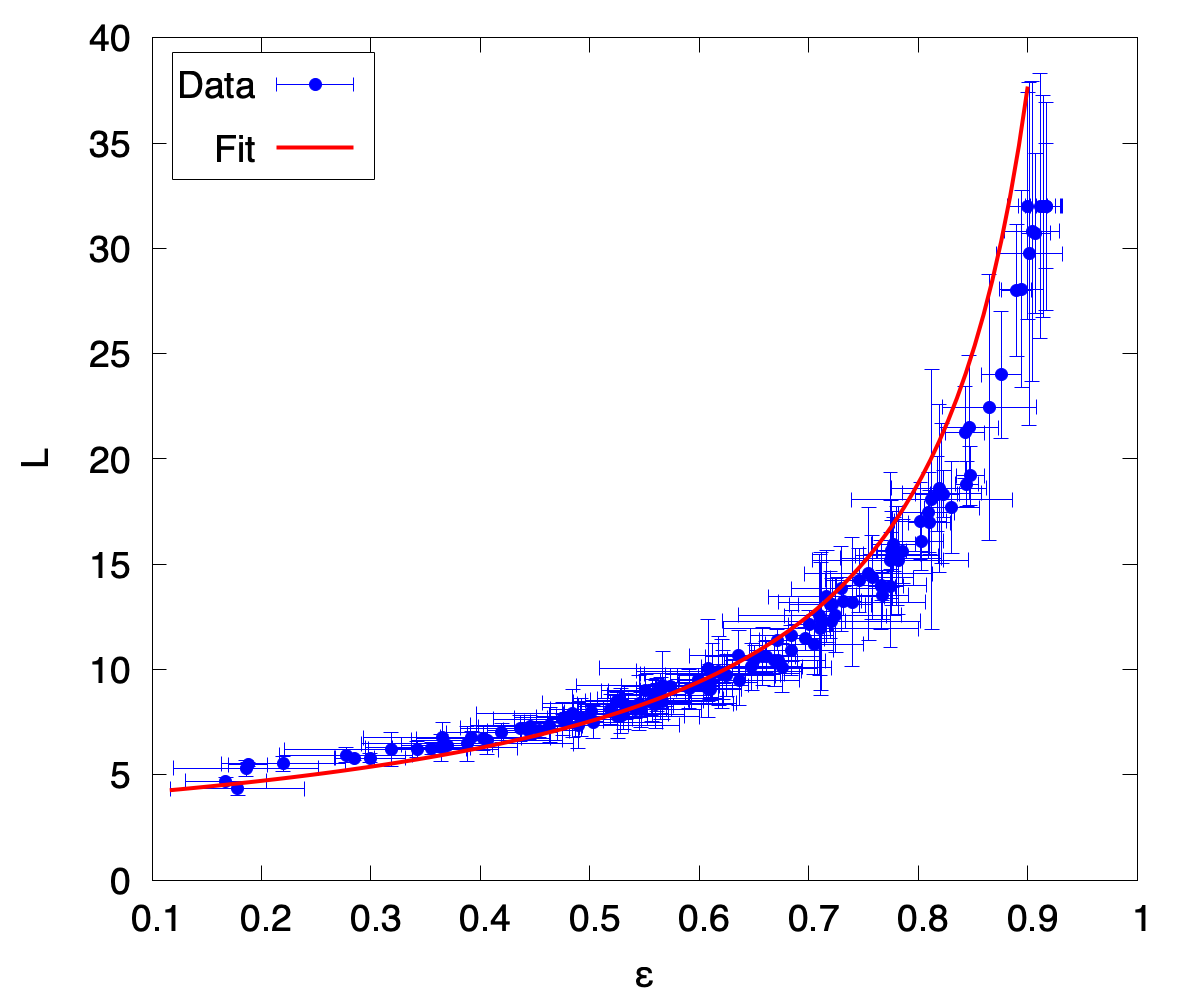}
	\caption{The relationship between the fitted parameters $L =
	\log(\lambda/1\text{ kpc})$ and $\epsilon$ for the 131-galaxy sample. The
	data points show a clear trend, which is well-described by the best-fit
	theoretical curve from Eq. \eqref{eq:l-r-rel}.}
	\label{fig:L-eps-fit}
\end{figure}
\begin{figure}[t]
	\centering
	\includegraphics[width=0.7\textwidth]{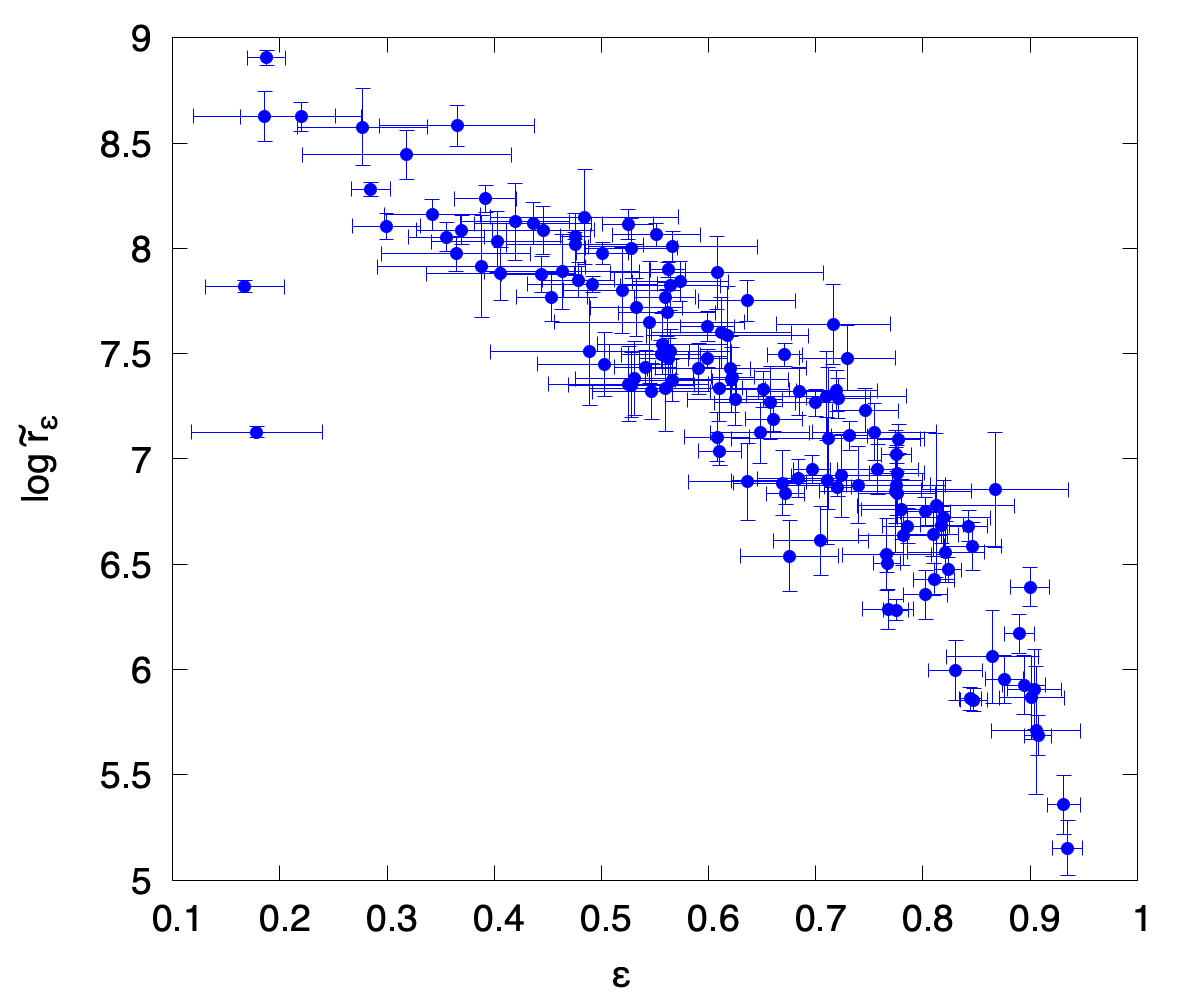}
	\caption{The calculated value of $\log\tilde{r}_\epsilon$ as a function of
	the fitted parameter $\epsilon$, showing a clear trend rather than a
	constant value.}
	\label{fig:r-eps}
\end{figure}
\begin{figure}[t]
    \centering
    \includegraphics[width=0.7\textwidth]{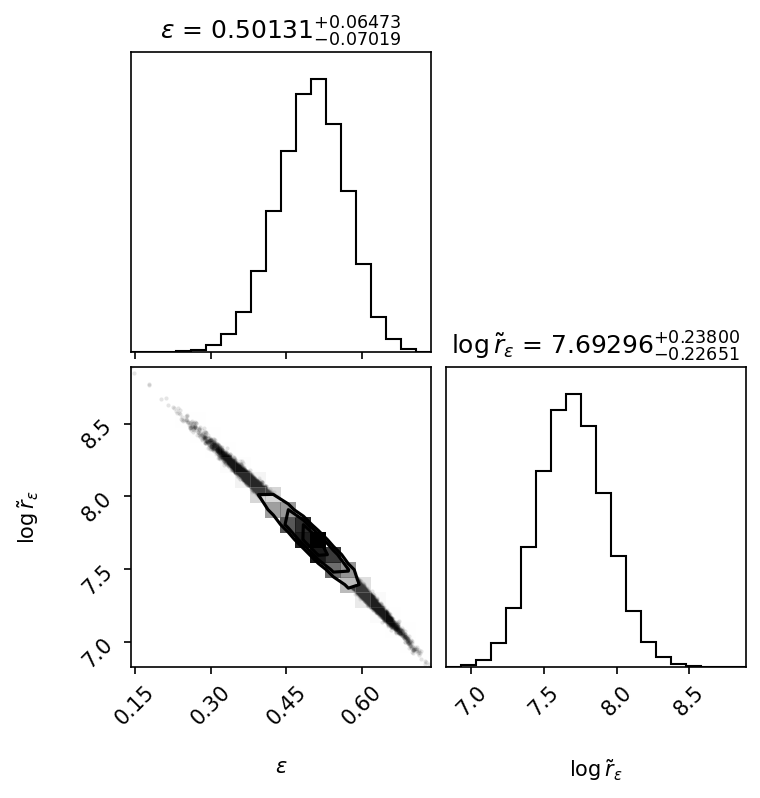}
    \caption{A corner plot showing the posterior distributions for the
    parameters $\epsilon$ and $\log\tilde{r}_\epsilon$ for NGC4157 ($\chi_\nu^2
    = 0.251$). The elongated contour illustrates the strong correlation between
    the two parameters.}
    \label{fig:corner}
\end{figure}

To further explore the parameter space of the Power-Law model, we analyze the
relationship between its two parameters. As discussed previously, a practical
difficulty emerges when fitting for $\lambda$ directly due to its large
magnitude. We therefore introduce the logarithmic scale parameter $L$, defined
as:
\begin{equation}
	L = \log\frac{\lambda}{1\text{ kpc}}.
\end{equation}
The relationship between $L$ and $\epsilon$ for our full sample of 131 galaxies
is shown in Figure \ref{fig:L-eps-fit}. The data are not randomly scattered but
follow a distinct trend, which is governed by the theoretical relation involving
the dimensionless strength parameter $\tilde{r}_\epsilon$:
\begin{equation}
	L = \frac{\log\tilde{r}_\epsilon}{2(1-\epsilon)}.
	\label{eq:l-r-rel}
\end{equation}
Fitting this function to the data yields a mean value of
$\log\tilde{r}_\epsilon = 7.53758 \pm 0.06731$.
This high degree of consistency motivates an investigation into whether
$\tilde{r}_\epsilon$ could be treated as a universal constant. However, an
explicit plot of $\log\tilde{r}_\epsilon$ against $\epsilon$ (Figure
\ref{fig:r-eps}) reveals that this is not the case; a clear trend suggests a
dependency between the two parameters. To visualize this dependency more
clearly, we perform a Bayesian analysis for a representative, well-fitting
galaxy (NGC4157), assuming flat priors. The resulting posterior distribution,
shown in the corner plot (Figure \ref{fig:corner}), confirms a strong
correlation between $\epsilon$ and $\log\tilde{r}_\epsilon$. Despite this
internal correlation, we were unable to find any significant correlation between
our model parameters and global galaxy properties available in the SPARC data,
such as effective radius or total mass.

\begin{figure}[t]
	\centering
	\includegraphics[width=0.7\textwidth]{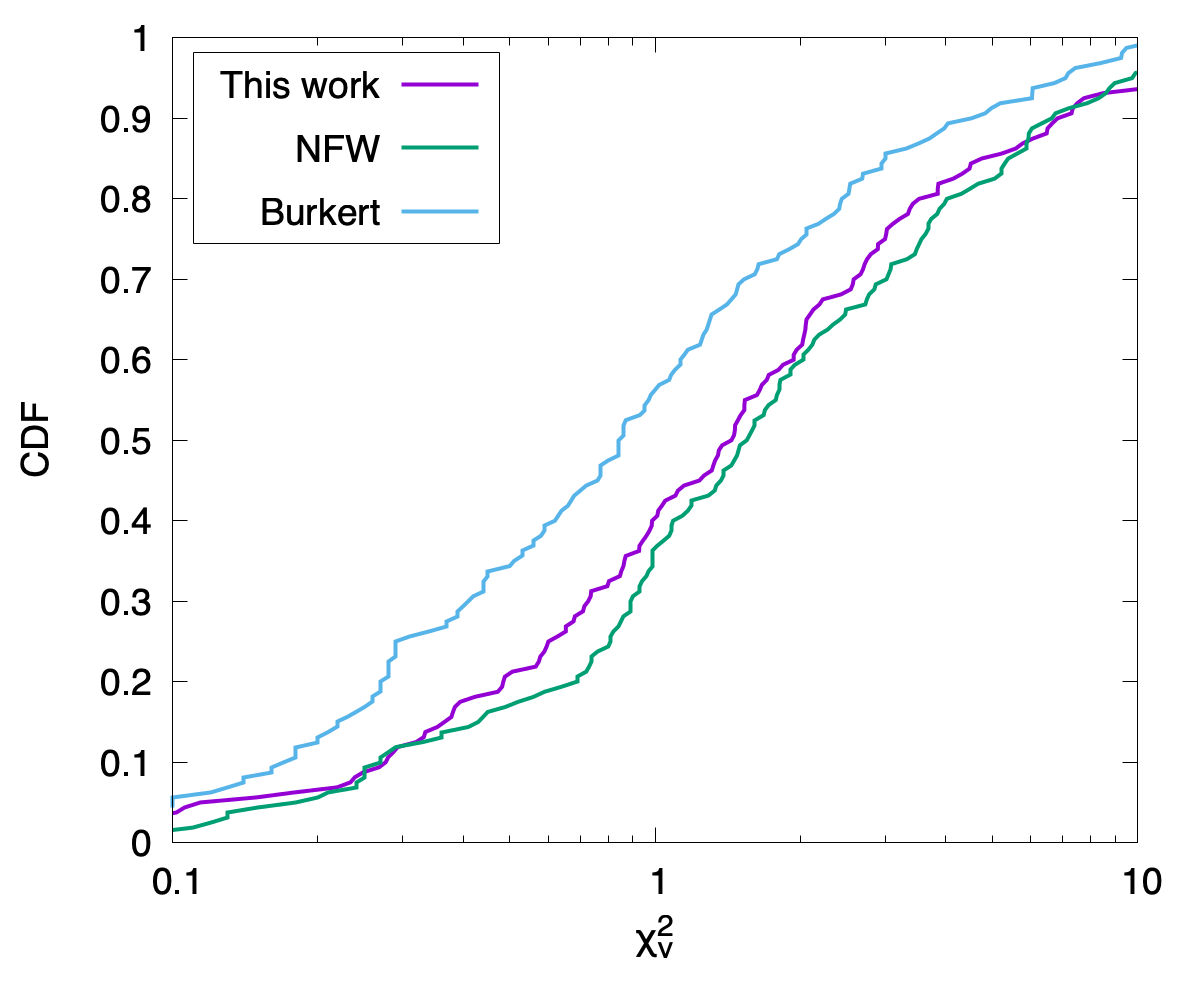}
	\caption{The cumulative distribution function (CDF) of reduced $\chi_\nu^2$
	values for our Power-Law model, the NFW profile, and the Burkert profile
	across the galaxy sample.}
	\label{fig:chi-cdf}
\end{figure}

To benchmark the performance of our model, we conduct a direct comparison with
two standard dark matter profiles: the cuspy Navarro-Frenk-White (NFW) profile
and the cored Burkert profile. For a fair comparison, we utilize the reduced
chi-squared values reported in \cite{li2020comprehensive}, which were also
obtained using flat priors. A qualitative overview is provided by the cumulative
distribution function (CDF) of the $\chi_\nu^2$ values for all three models,
presented in Figure \ref{fig:chi-cdf}. The CDF indicates that our model
generally provides a better fit than the NFW model but is significantly
outperformed by the empirical Burkert profile. This result is expected, as our
model's cored-like nature gives it an advantage over the cuspy NFW profile,
while the Burkert model, being empirically constructed, possesses greater
flexibility.

To provide a more quantitative assessment, we employ two simple statistical
tests, as a full comparison using metrics like the Bayesian Information
Criterion (BIC) is not feasible with the available data. First, an analysis of
the ratio of $\chi_\nu^2$ values reveals a median of 0.96 for our model relative
to NFW, implying our model provides a better fit for half of the
galaxies in the sample. In contrast, the median ratio relative to the Burkert
profile is 1.55, indicating a typical fit that is $\approx55\%$ worse. Second, we examine
the fraction of "good fits," defined as $\chi_\nu^2 < 1.5$. Our model achieves
this for $52.5\%$ of the galaxies, compared to $48.8\%$ for NFW and $69.5\%$ for
Burkert. Both tests confirm that our simple, two-parameter theoretical model is
not only competitive with but sometimes better than the standard NFW profile,
reinforcing its physical relevance.

A crucial theoretical point must be addressed regarding the Power-Law solution.
For the fitted range of $\epsilon \in (0, 1)$, the dark matter term in the
metric function $f(r)$ in Eq. \eqref{eq:met-exp} diverges as $r \to \infty$.
This implies that the spacetime is not asymptotically flat, and therefore this
metric cannot be a valid global solution for an isolated galaxy. This limitation
suggests that the Power-Law metric should be interpreted as an \textit{effective}
description, valid only within the radial extent of the dark matter halo where
it provides a good approximation to the local spacetime geometry.

In summary, our analysis indicates that the Logarithmic metric successfully
captures the asymptotic behavior of the halo at large radii, while the Power-Law
metric effectively models the inner halo structure, whose properties align more
closely with a cored density profile. The fact that these two distinct behaviors
emerge from a single, simple fluid model parameterized by $\epsilon$ provides a
compelling, if phenomenological, framework for understanding the distribution of
dark matter.

\section{Conclusions}
In this work, we have investigated the viability of a dark matter model derived
from a perfect fluid with a simple, barotropic equation of state within the
framework of General Relativity. The primary goal was to ascertain whether such
a minimalist and theoretically-grounded model, developed without recourse to
empirical density profiles, could account for the observed rotation curves of
galaxies. Our analysis, based on a comparison with high-quality data from the
SPARC database, demonstrates that this approach is not only viable but also
offers valuable physical insights.

We have shown that the two solutions arising from this framework: the Power-Law
and Logarithmic metrics act as complementary descriptions of the dark matter
halo. The Logarithmic solution effectively reproduces the asymptotic behavior of
the rotation curve at large radii, a region where many established models
converge. Concurrently, the Power-Law solution provides an excellent description
of the inner halo, successfully modeling the initial rise and subsequent
flattening of the velocity profile. The model's inherent preference for a
gradual rise in velocity indicates that it generates a density profile more akin
to a cored halo than a cuspy one. This finding is particularly relevant to the
ongoing \textit{cusp-vs-core} debate, positioning our model as a potential theoretical
basis for the observationally favored cored profiles.

A quantitative comparison of the goodness-of-fit reveals that our model performs
favorably against the Navarro-Frenk-White (NFW) profile. Statistical tests show
that our model yields a lower reduced chi-squared value than the NFW profile for
half of the galaxies in the sample. As anticipated, the more flexible,
three-parameter empirical Burkert profile provides a better overall fit to the
data. The key result is that our simple, analytically-derived model is not only
physically motivated but is also statistically competitive with the standard
cuspy NFW model.

This work opens several promising avenues for future research. While this paper
has successfully demonstrated the model's viability on a select sample of
galaxies, it should be viewed as a foundational proof of concept. A crucial next
step is to perform a more exhaustive statistical analysis across a much larger
sample of rotation curves. This would
rigorously test the model's universality and explore any potential correlations
between the $\epsilon$ parameter and galaxy properties like type, or
environment. Beyond this expanded empirical validation, other theoretical
extensions are warranted. A natural progression would be to move beyond a
constant parameter $\epsilon$ and explore a radially dependent equation of
state, where $\epsilon$ becomes a function $\epsilon(r)$. Such a modification
could potentially unify the two solutions into a single, seamless description.
Furthermore, the model could be enriched by considering interactions between
this dark matter fluid and other physical fields, such as the electromagnetic
field \cite{rogatko2025influence}. Ultimately, the framework presented here
offers a robust and theoretically elegant starting point for further exploration
into the fundamental properties of dark matter.

\section*{Acknowledgements}
The author wishes to express sincere gratitude to S. Płatek for the
invaluable discussions and crucial insights that helped to conceptualize and
shape the direction of this work.

This research has made use of the SPARC (Spitzer Photometry and Accurate Rotation
Curves) database. We are grateful to the creators and maintainers of SPARC for
making this comprehensive and high-quality dataset publicly available, as it was
fundamental to the analysis presented in this paper.
\bibliography{refs}
\end{document}